# Comments on "Limits on Dark Matter Using Ancient Mica"

The search for WIMP-recoil tracks in mica using Atomic Force Microscopy [1] promises important improvements on current exclusion limits. For this to be fulfilled, background sources that could mimic the WIMP-recoil signature must be rejected. In the case of naturally occurring α-recoil tracks, the authors rely on the longer expected depth of their etch-pits, since they result from the sum of 8 (6) successive and spatially-connected recoils of nuclei in the $^{238}$U ($^{232}$Th) chains. An isolated α-recoil would be essentially indistinguishable from a typical WIMP-recoil: for instance, the ~ 70 keV $^{234}$Th α-recoil has a range P ~ 290 ± 55 ± 45 Å in mica (the errors are longitudinal and lateral straggling respectively), while the recoil induced by a 100 (500) GeV/$c^2$ WIMP scattering from a potassium nucleus has an average kinetic energy of ~ 30 (42) keV, with P ~ 270 ± 90 ± 70 (365 ± 120 ± 90) Å. The larger total stopping power $S_t$ for the $^{234}$Th α-recoil ($S_t$ ~ 13.6 vs ~3.8 GeV cm$^2$/g) guarantees a comparatively higher pit-revealing efficiency in the etching process, and a larger resistance to thermal annealing. The probability that all the consecutive α-recoil damage tracks retrace each other's trajectories imitating the WIMP signature is tiny. This has prompted the authors to claim that their method is at present not background-limited.

I would like to suggest a possible weakness in this reasoning. The $^{238}$U chain proceeds as $^{238}$U $\xrightarrow{\alpha}$ $^{234}$Th $\xrightarrow{\beta^-}$ $^{234}$Pa $\xrightarrow{\beta^-}$ $^{234}$U $\xrightarrow{\alpha}$ $^{230}$Th... with α-emission half-lives $T^{238}_{1/2} = 4.4 \cdot 10^9$ y, $T^{234}_{1/2} = 2.4 \cdot 10^5$ y (the very fast β$^-$ decays produce no observable recoil tracks). The long $T^{234}_{1/2}$ results in some $^{234}$Th recoils not being accompanied by a second one, thereby mimicking a WIMP recoil. These single α-recoils are not negligible in this search: for a mica age A $\lesssim 10^9$ y, one can approximate that the $^{238}$U decay is taking place at a constant rate $R_{238}$ ~ N $\lambda_{238}$, where $\lambda_{238} = (\ln 2)/T^{238}_{1/2}$ and N is the number of $^{238}$U atoms in the sample. The number of $^{234}$U atoms (i.e., the number of single α-recoils) in equilibrium (A >> $T^{234}_{1/2}$) is n ≈ $R_{238}/\lambda_{234}$ = N ($T^{234}_{1/2}/T^{238}_{1/2}$) = 5.4·10$^{-5}$ N. The authors estimate the concentration of U and Th to be a < 10$^{-10}$ atom fraction, 10$^{-11}$ in the purest mica. This is N > 10$^{-11}$ x 9·10$^{22}$ $^{238}$U atoms / cm$^3$, or a lower limit of n > 5.4·10$^{-5}$ N = 4.9·10$^7$ $^{234}$U atoms / cm$^3$, n also being the density of latent "fake" WIMP tracks. The surface density of cleavage-crossing pits revealed after an etching time t of a cleaved surface is approximately given by [2] $\rho = \frac{1}{2} n\, P \cos^2\theta_c$, where $\theta_c$ is the critical etching angle ($\sin\theta_c = V_\perp / V_t$). For etching in 25° HF, $V_t \approx 0.012 \cdot S_n$ μm / h with the nuclear stopping power $S_n$ in GeV cm$^2$ / g (here $S_n = 12.4$) and $V_\perp \approx 0.027$ μm / h [3]. These velocities are conservative in view of the authors' updated values [1] and their demonstration of the non-negligible role of the electronic stopping power in creating solid state damage [4]. For t= 1 h, the surface density of cleavage-crossing shallow "fake" WIMP pits is then $\rho \gtrsim 70 / $ cm$^2$ for the purest mica. The small $\theta_c$ for this $\alpha$-recoil allows a large fraction of trajectories slant to the cleaved surface to be etched, forming pits in the $\lesssim 100$ Å depth region ($\gtrsim 4$ % falling in the summed etched depth 40-64 Å).

In view of this, the AFM mica search (having covered 1 / 1240 cm$^2$) might soon run into this competitive background, not to be taken for a WIMP signal. If this is so and a reasonable rejection method cannot be devised, AFM mica searches will be at best just competitive with current underground experiments. A direct comparison of n with the volume density of stored WIMP tracks leads to the same conclusion. A technologically less involved option would be to search for a possible large depth ($\gtrsim 500$ Å) tail of the WIMP pit distribution [4], but this is subordinate to the importance of the $\alpha$-recoil distribution in that region.

This caveat is unfortunate, since the AFM mica technique could have been exploited for more than WIMP searches. Calculations similar to those in [4] show that recoils from stellar collapse neutrinos and $^8$B solar neutrinos would induce etchable tracks with $\rho \gtrsim 0.03 / $cm$^2$ in the $\lesssim 100$ Å depth region, at fluxes predicted by standard models of the galaxy and Sun. Competitive experimental limits on the stellar collapse rate in our galaxy could have been obtained with O(1) cm$^2$ areas. Similarly, the large historical variations in the temperature of the solar core and hence in the $^8$B neutrino flux that some solar models predict could have been ruled out.

However, there are other areas where AFM mica searches can yield new limits. Silk and Stebbins have recently suggested that "clumps" of cold dark matter might abound in our galaxy [5]. For some string-seeded models of the early universe, the cores of these clumps have densities ~$10^9$ larger than the local halo density. In favorable cases, the rapid crossing of these clumps through the solar system (taking O(1) y) occurs every few tens of My, with possible effects on the biosphere [6]. The expected rate of WIMP interaction with micas of age ~ 1 Gy would then be up to two orders of magnitude larger than presently expected, and mica limits on this interesting possibility will soon be available.

**Juan I. Collar**

**Department of Physics and Astronomy**

**University of South Carolina**

**Columbia, SC 29208**